\documentclass[10pt,a4paper,twocolumn,aps,pra,showpacs,superscriptaddress]{revtex4-1}
\pdfoutput=1
\usepackage[utf8x]{inputenc}
\usepackage{ucs}
\usepackage{amsmath}
\usepackage{amsfonts}
\usepackage{amssymb}
\usepackage{makeidx}
\usepackage{booktabs}
\usepackage{natbib}
\usepackage{lipsum}

\usepackage{color}
\definecolor{light-gray}{gray}{0.55}

\usepackage{microtype}

\usepackage{graphicx}

\renewcommand{\dag}{^{\dagger}}
\newcommand{\exv}[1]{ \langle #1 \rangle }

\newcommand{\bra}[1]{ \langle #1 \rvert }
\newcommand{\ket}[1]{ \lvert #1 \rangle}

\newcommand{\tr}[2][]{\text{Tr}_{ #1 } ( #2 )}

\newcommand{\pfrac}[2]{\frac{\partial #1}{\partial #2}}

\begin{document}

\begin{abstract}
A Lagrangian formalism is used to derive the Hamiltonian for a $\lambda$/4-resonator shunted by a current-biased Josephson junction. The eigenstates and the quantum dynamics of the system are analyzed numerically, and we show that this quantum system can function as an efficient detector of weak incident microwave fields.
\end{abstract}

\date{\today}
\author{Christian Kraglund Andersen}
\thanks{E-mail: ctc@phys.au.dk}
\affiliation{Department of Physics and Astronomy, Aarhus University, DK-8000 Aarhus, Denmark}
\author{Gregor Oelsner}
\affiliation{Leibniz Institute of Photonic Technology, P.O. Box 100239, D-07702 Jena, Germany}
\author{Evgeni Il'ichev}
\affiliation{Leibniz Institute of Photonic Technology, P.O. Box 100239, D-07702 Jena, Germany}
\affiliation{Novosibirsk State Technical University, 20 K. Marx Ave., 630092 Novosibirsk, Russia}
\author{Klaus Mølmer}
\affiliation{Department of Physics and Astronomy, Aarhus University, DK-8000 Aarhus, Denmark}

\title{Quantized resonator field coupled to a current-biased Josephson junction in circuit QED}


\maketitle

\section{Introduction}

During the last decades, quantum optical system behavior has been implemented in solid state systems using superconducting Josephson junctions and transmission wave guides \cite{Wallraff:2004oh,you2011atomic,RevModPhys.73.357,vion2002manipulating}. In these studies, electric circuit resonators take the place of cavities while the Josephson junction non-linearity gives rise to effective few-level systems. The emerging field of circuit quantum electrodynamics (cQED) offers promising perspectives for quantum information processing with manufactured, scalable systems \cite{PhysRevA.75.032329,zakosarenko2007realization,dicarlo2009demonstration}.

Josephson junctions are used in a vast number of experiments exploring the macroscopic quantum nature of the junction phase variable. This allows studies of macroscopic quantum tunneling and of microwave driving among the quantized levels in the junction \cite{PhysRevLett.55.1543,PhysRevB.35.4682,PhysRevLett.89.117901,Yu02}. The tunneling mechanism of the Josephson junctions is also used as a readout mechanism of metastable qubits in these studies.
Josephson junctions are furthermore used in Josephson parametric amplifiers (JPA), where the non-linearity of the junction allows low-noise amplification of weak microwave fields \cite{PhysRevA.39.2519,yamamoto2008flux,castellanos2008amplification,bergeal2010phase,PhysRevLett.107.113601}. Today this technique has led to quantum limited detectors in the microwave regime \cite{RevModPhys.82.1155,PhysRevA.86.032106}, but the detection of single microwave photons is still a major challenge.

\begin{figure}[t]
\includegraphics[width=0.95\columnwidth]{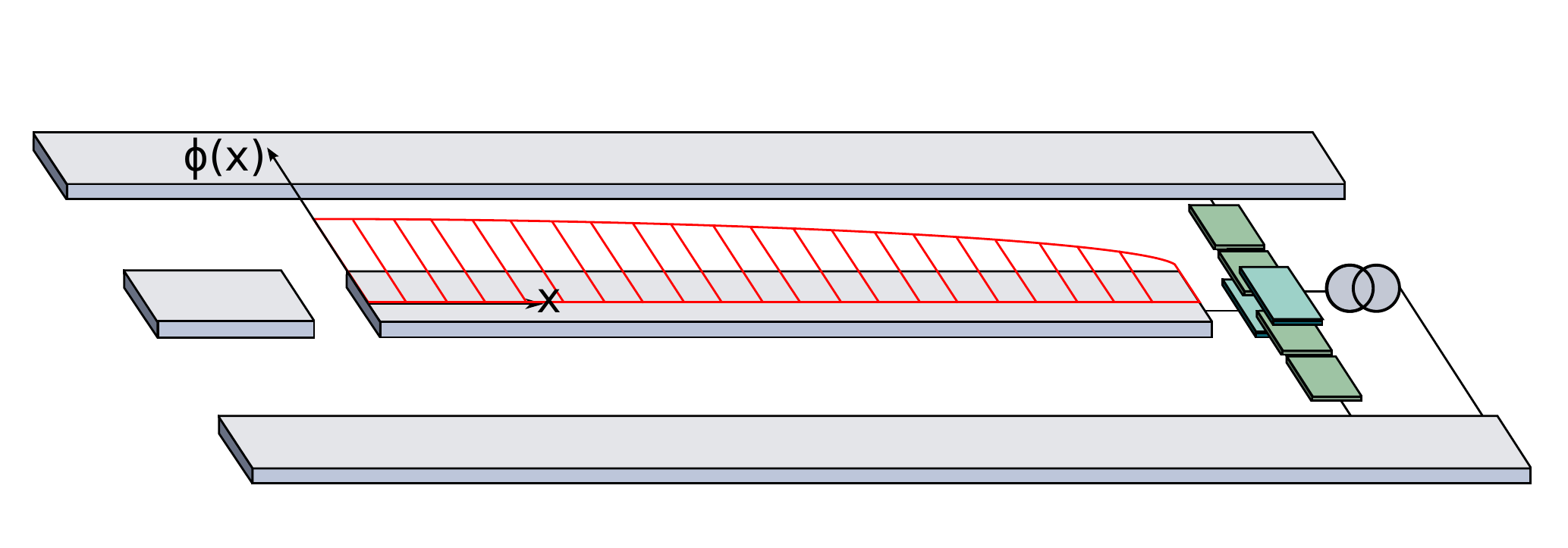}
\includegraphics[width=0.95\columnwidth]{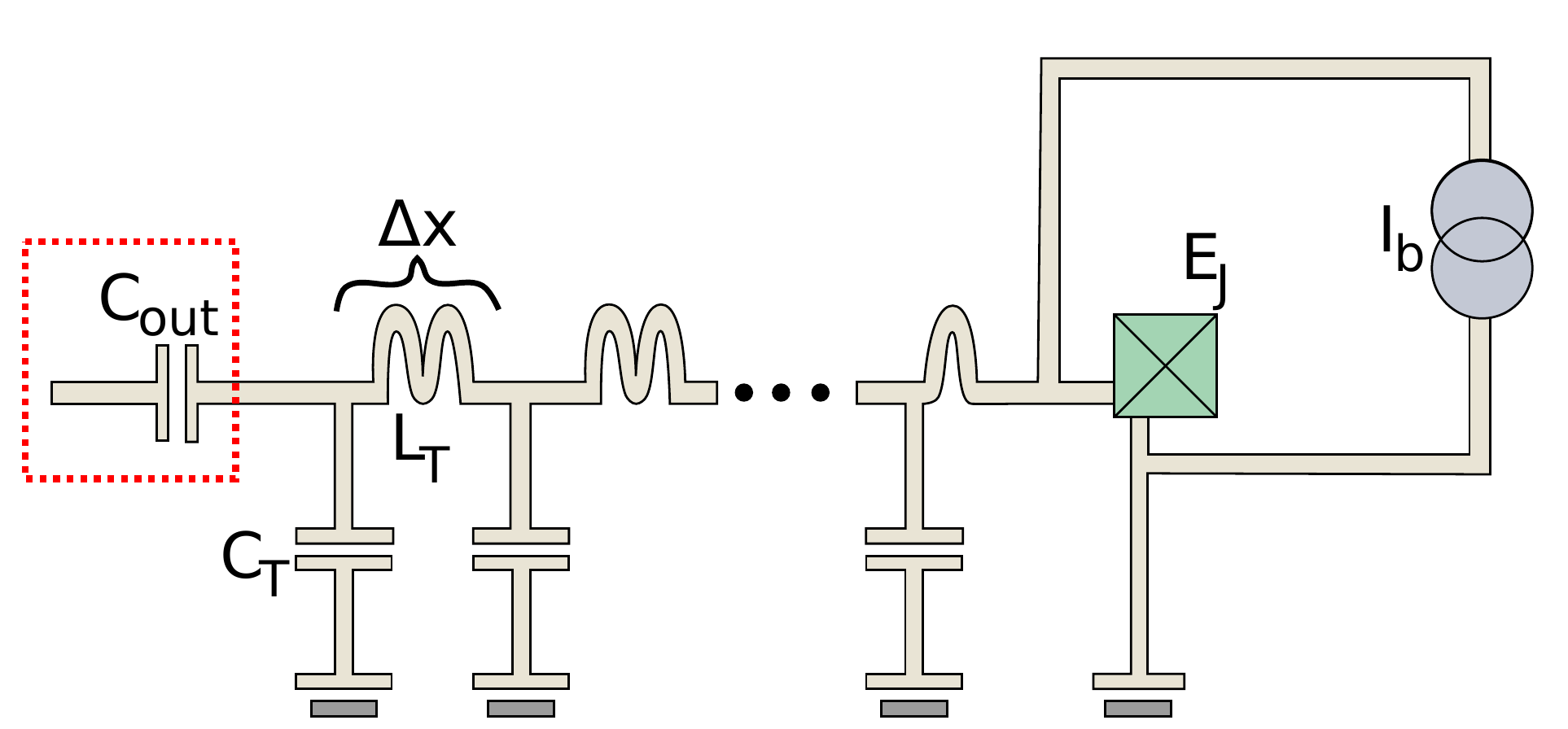}
\caption{(Color online) A schematic (top) and a circuit diagram (bottom) of the system. In the upper picture the fundamental normal mode of the flux is sketched. In the circuit diagram, the transmission waveguide resonator is described as a chain of capacitors and inductors. The resonator is shunted by a current biased Josephson junction. In the dashed (red) box, we couple capacitively to fields outside of the device.} \label{fig:circuit}
\end{figure}

Recent works using current biased Josephson Junctions (CBJJ) have made both experimental \cite{PhysRevLett.107.217401} and theoretical \cite{PhysRevLett.102.173602,PhysRevA.84.063834,PhysRevB.86.174506} progress towards a single microwave photon detector. The aim of this work is to contribute to these developments by studying the response of the CBJJ coupled to a $\lambda$/4-resonator in the few photon regime (see Fig. \ref{fig:circuit}). The general idea is to use the device both as an amplifier and as a detector, sensitive to a single or a few quanta in the resonator through a classical measurable response in the form of a voltage switch over the Josephson junction.

In Sec. \ref{sec:coupling}, we discuss the general framework of cQED and outline the challenges in coupling a junction to a resonator. Section \ref{sec:derivation} is devoted to the derivation of the Hamiltonian for the system of a $\lambda$/4 resonator shunted by a CBJJ. We present an eigenvalue analysis of this Hamiltonian in Sec. \ref{sec:spectral} and a time-dependent analysis in Sec. \ref{sec:time}. Section \ref{sec:conclusion} concludes the paper.

\section{Coupling in circuit QED}
\label{sec:coupling}

A microwave transmission wave guide is conveniently described as an infinite series of $LC$ circuits \cite{PhysRevA.29.1419}. The coupling of the circuits leads to the identification of oscillator eigenmodes, in which one can quantize the system and obtain the usual Hamiltonian
\begin{align}
H_T = \sum_j \hbar \omega_j \, a_j\dag a_j
\end{align}
with $\omega_j$ being the angular frequency and $a_j$ ($a_j\dag$) being the annihilation (creation) operator of photons in the $j$'th mode. When appropriate (see Sec. \ref{sec:siglemode}) a single mode approximation can be made, which reduces the Hamiltonian to $H_{T,s} = \hbar \omega \, a\dag a$.

A CBJJ is described by the effective Hamiltonian \cite{devoret2004}
\begin{align}
H_{JJ} = - \frac{\hbar^2}{2M} \pfrac{^2}{\varphi^2} - E_J (\cos \varphi + I\varphi) \label{eq:H_jj}
\end{align}
which describes the phase as the position of a particle moving in or trapped in a well of a washboard potential with effective mass $M = C_J/(2e)^2$, where $C_J$ is the Josephson capacity and $E_J$ is the Josephson energy.

The junction provides an easily accessible readout mechanism as the voltage across the junction will increase when the particle goes from being trapped in a well to running down the potential. Our goal is to provide a consistent theoretical description of the coupling between the motion of the phase particle and the microwave resonator field, i.e. to derive a Hamiltonian in the form
\begin{align}
H = H_{T,s} + H_{JJ} + H_I
\end{align}
where $H_I$ will contain coupling and interaction terms. Note that we want a full description of the junction degree of freedom. Previous works  \cite{you2011atomic,devoret2004,RevModPhys.85.623,omelyanchouk2010quantum,niemczyk2010circuit,PhysRevA.80.032109} on the matter reduces the junction to be either described by a simple two-level system or use the junction in combination with other junctions to create a SQUID, which can also be viewed as a two-level system.

In the well-known Rabi-model of a two-level system coupled to a field the interaction Hamiltonian reduces to $H_I = g(a + a\dag)\sigma_x$, where $\sigma_x$ is the Pauli $x$-matrix. The Rabi-model can be solved in the rotating wave-approximation, and also in general  \cite{PhysRevLett.107.100401,PhysRevA.86.015803}. However, since we here want to describe the full behavior of the Josephson junction, including the switching dynamics and the dissipation of the junction, we cannot in general apply the simple Hamiltonian of the Rabi-model.

\section{Derivation of the Hamiltonian}
\label{sec:derivation}

In order to obtain the Hamiltonian for the system, we consider the corresponding classical system for which we can directly write up the Lagrangian. With the Lagrangian at hand we can identify the canonical variables and perform a canonical quantization and a Legendre transform to obtain the quantum mechanical Hamiltonian. The approach followed here is similar to the approaches of  \cite{PhysRevB.74.224506,eichler2013controlling,devoret1995quantum,PhysRevA.29.1419}.

An electrical circuit can be described as a network of electrical elements, e.g., capacitors and inductors, known as branches. We introduce the node variables, $\phi_n$ and $q_n$, associated with every node of the electrical circuit diagram. The flux variable $\phi_n$ is defined as the time-integral of the voltage measured along a path of branches, called the spanning tree, connecting the node to the ground. Branches not included in these paths are called closure branches. The equation of motion for node variables, will in general depend on the chosen topology of the spanning tree \cite{devoret1995quantum,PhysRevA.29.1419}.

In Fig. \ref{fig:circuit}, we show a lumped element representation of our system. We will choose the bias line to be a closure branch of the system, while the rest constructs the spanning tree. The resonator is here depicted as a series of $n$ $LC$ circuits, which in the limit $n\rightarrow\infty$ will give an appropriate description of the resonator.

This now allows us to write the equation of motion for each node of the resonator, except the end node,
\begin{align}
C_T \Delta x \, \ddot{\phi}_j = \frac{\phi_{j+1} -\phi_{j-1}}{L_T \Delta x} \qquad \text{for } 1 \leq j \leq n-1, \label{eq:eom_j}
\end{align}
with $\Delta x = d/n$, $d$ being the length of the resonator and $C_T$ and $L_T$ being the capacitance and inductance per length of the resonator. Taking the continuum limit of $\Delta x \rightarrow 0$, our sequence of discretized flux variables become a function of $x$, $\phi_j \rightarrow \phi(x)$, and Eq. \eqref{eq:eom_j} reduces to the wave equation
\begin{align}
\frac{1}{C_TL_T} \partial_x^2 \phi(x) - \partial_t^2 \phi(x) = 0.
\end{align}

The end point of the circuit is shunted with a bias current, $I_b$, which we model as a high inductance line, with the inductance $L_S$, pre-charged with a large flux, $\tilde{\Phi}_S$, such that $\tilde{\Phi}_S /L_S = I_b$. The equation of motion is then
\begin{align}
C_J \, \ddot{\phi}_n \underset{\phantom{L_S \rightarrow \infty}}{=}& \frac{\phi_n - \phi_{n-1}}{L_T\Delta x}- \frac{2e}{\hbar} E_J \sin{\frac{2e}{\hbar} \phi_n} + \frac{\tilde{\Phi}_S - \phi_n}{L_s} \\
\underset{L_S \rightarrow \infty}{=}& \frac{\phi_n - \phi_{n-1}}{L_T\Delta x}-\frac{2e}{\hbar} E_J \sin{\frac{2e}{\hbar} \phi_n} + I_b.
\end{align}

We are now able to write the proper Lagrangian for the system, such that the Euler-Lagrange equations give the above equations of motion:

\begin{align}
\mathcal{L} =& \int_0^d dx \bigg\{ \frac{C_T}{2} \big( \partial_t \phi(x) \big)^2 - \frac{1}{2L_T} \big( \partial_x \phi(x) )^2 \bigg\} \nonumber \\
&+ \frac{C_J \big( \partial_t \phi(d) \big)^2}{2} + E_J\Big( \cos {\frac{2e}{\hbar} \phi(d)} + I\, \frac{2e}{\hbar} \phi(d)  \Big), \label{eq:lagran}
\end{align}
with $I = I_b / I_c$, where the critical current is defined as $I_c = \frac{2e}{\hbar} E_J$.

The phase across the Josephson junction, $\phi_J$, is given as a function of the bias current,
\begin{align}
\frac{2e}{\hbar} \phi_J = \sin^{-1} I,
\end{align}
which will also contribute with a predefined flux in the transmission resonator. If we neglect contributions from the Josephson capacitance, $C_J$, the Euler-Lagrange equation at $x=d$ yields
\begin{align}
\frac{1}{L_T} \partial_x \phi(d) =    \frac{2e}{\hbar} E_j \Big( \sin {\frac{2e}{\hbar} \phi(d)} + I  \Big). \label{eq:lineu}
\end{align}

Generally the flux bias will not be constant, but it leads us to the ansatz for solutions of the equation of motion given by
\begin{align}
\phi(x) = \sum_j \phi_j \, \cos {k_j x} + \phi_0 \label{eq:ansatz}.
\end{align}
In writing Eq. \eqref{eq:ansatz}, we have assumed that there is no incident field at the capacitor, $C_{out}$, leading to the open boundary condition $\partial_x \phi = 0$ at $x=0$. The open boundary condition is equivalent to the assumption of a vanishing current, while the  time-derivative of $\phi(0)$ yields the voltage at $C_{out}$,  determined by the field inside the resonator. The values of $k$ in Eq. \eqref{eq:ansatz} must be chosen to match the boundary condition following from the linearized Euler-Lagrange equation at $x=d$ [Eq. \eqref{eq:lineu}].

Using the steady state result for the junction phase $\phi_0 = \phi_J$, with the approximation that the phase difference between the phase across the junction and the steady state phase is small, that is $\sum \phi_j \cos k_jd \ll 1$, we can derive the following approximate identity
\begin{align}
& \bigg(  \sin {\frac{2e}{\hbar} \phi(d)} + I \bigg) = \frac{2e}{\hbar} \sum_j \phi_j \cos k_jd \, \cos\frac{2e}{\hbar}\phi_0,
\end{align}
and we obtain the linearised equation for each independent mode
\begin{align}
k_jd \tan k_jd = \frac{L_T d}{L_J} \cos \frac{2e}{\hbar}\phi_J, \label{eq:el-k}
\end{align}
with $L_J = (\hbar/2e)^2 / E_J$. This equation can be solved numerically or approximated by
\begin{align}
k_j d = \frac{\pi (1+ 2j) }{2\big(1 + \frac{L_J}{L_T d \cos \phi_J}\big)},
\end{align}
valid for $L_J \ll L_Td \cos \frac{2e}{\hbar} \phi_J$. We recall that our approximate solutions are only valid when neglecting the Josephson capacitance, $C_J$. In the following we shall reinstate a contribution from $C_J$ and evaluate its influence on the modes defined in \eqref{eq:ansatz}.

\subsection{Single-mode approximation} \label{sec:siglemode}

Having Eq. \eqref{eq:ansatz} as a solution for the normal modes we can choose to look at a single-mode field
\begin{align}
\phi(x) = \phi \cos kx + \phi_0
\end{align}
and substitute this solution into the Lagrangian
\begin{align}
\mathcal{L} =& \, \dot{\phi}^2 \Big( \int_0^d \frac{C_T \cos^2 kx}{2} dx + \frac{C_J \cos^2 kd}{2} \Big)\nonumber\\& + \dot{\phi}_0^2 \frac{C_T d + C_J}{2} \nonumber\\&+ \dot{\phi} \dot{\phi}_0 \Big(\int_0^d C_T \cos kx \, dx + C_J \cos kd \Big)\nonumber\\&- \phi^2 \int_0^d \frac{k^2 \sin^2 kx}{2L_T} dx \nonumber\\&+ E_J \Big( \cos \frac{2e}{\hbar} (\phi \cos kd + \phi_0) \nonumber\\&\phantom{+ E_J \Big(}+ \frac{2e}{\hbar} I (\phi \cos kd + \phi_0) \Big).
\end{align}
Next, we expand the $\cos$-term of the potential as $\cos(A+B) = \cos A \cos B - \sin A \sin B$ followed by an expansion to fourth order of $(\frac{2e}{\hbar}\phi \cos kd )$ allowing also for Kerr-effects in the device. We can reduce the expressions in the Lagrangian significantly by introduction of the quantities
\begin{align}
C_E &= \frac{C_Td}{2} \Big(1 + \frac{\sin 2kd}{2kd} \Big) + C_J \cos^2 kd \\
C_0 &= C_Td + C_J \\
C_c &= C_Td\,\frac{\sin kd }{kd} + C_J \cos kd \\
L_E^{-1} &= \frac{(kd)^2}{2L_Td} \Big( 1 - \frac{\sin 2kd}{2kd} \Big).
\end{align}
This constitutes the Lagrangian
\begin{align}
\mathcal{L} =& \, \dot{\phi}^2 \frac{C_E}{2} - \phi^2 \frac{1}{2L_E}  + \dot{\phi} \dot{\phi}_0 C_c  \nonumber\\&+ \dot{\phi}_0^2 \frac{C_0}{2}+ E_J\Big( \cos \frac{2e}{\hbar} \phi_0 + \frac{2e}{\hbar} I \phi_0 \Big) \nonumber\\& - E_J \Big( \frac{(2e)^2\phi^2 \cos^2 kd}{2\hbar^2} - \frac{(2e)^4\phi^4 \cos^4 kd}{24\hbar^4} \Big) \cos \frac{2e}{\hbar}\phi_0 \nonumber\\&-E_J\frac{2e}{\hbar} \phi \cos kd \Big( \sin \frac{2e}{\hbar}(\phi_0-\phi_J) \cos \frac{2e}{\hbar}\phi_J \nonumber\\&\phantom{-\frac{2e}{\hbar}E_J \phi \cos kd \Big(}- \frac{(2e)^2\phi^2 \cos^2 kd}{6\hbar^2} \sin \frac{2e}{\hbar}\phi_0 \Big),
\end{align}
from which we will derive the Hamiltonian. We introduce the conjugate variables to $\phi$ and $\phi_0$,
\begin{align}
q =& \pfrac{\mathcal{L}}{\dot{\phi}} = C_E \dot{\phi} + C_c \dot{\phi}_0 \\
q_0 =& \pfrac{\mathcal{L}}{\dot{\phi}_0} = C_0 \dot{\phi}_0 + C_c \dot{\phi},
\end{align}
and we perform a Legendre transformation to get the Hamiltonian
\begin{align}
H =&\, \frac{q_0^2}{2(C_0 - C_c^2/C_E)} - E_J \Big( \cos \frac{2e}{\hbar} \phi_0 + \frac{2e}{\hbar} I \phi_0 \Big) \nonumber\\&- \frac{C_c}{C_0C_E - C_c^2}qq_0 + \frac{q^2}{2 ( C_E - C_c^2/C_0)}  \nonumber\\& + \frac{\phi^2}{2L_E} +  E_J \frac{(2e)^2 \cos^2 kd}{2 \hbar^2}  \phi^2\, \cos \frac{2e}{\hbar}\phi_0 \nonumber\\&-E_J\frac{(2e)^4\cos^4 kd}{24 \hbar^4} \phi^4 \, \cos\frac{2e}{\hbar} \phi_0 \nonumber\\&-E_J\frac{2e}{\hbar} \phi \cos kd \Big( \sin \frac{2e}{\hbar}(\phi_0-\phi_J) \cos \frac{2e}{\hbar}\phi_J \nonumber\\&\phantom{-\frac{2e}{\hbar}E_J \phi \cos kd \Big(}- \frac{(2e)^2\phi^2 \cos^2 kd}{6\hbar^2} \sin \frac{2e}{\hbar}\phi_0 \Big). \label{eq:Hfull}
\end{align}

In the quantum regime the resonator operators $q$ and $\phi$ satisfy the canonical commutation relation $[\phi, q] = -i\hbar$, which allows us to introduce the ladder operator $a$ ($a\dag$) that annihilates (creates) a photon in the normal mode of the resonator. We write
\begin{align}
\phi &= i\sqrt{\frac{\hbar \omega L_E}{2}} (a - a\dag) \label{eq:defphi}\\
q &= \sqrt{\frac{\hbar}{2 \omega L_E}} (a + a\dag) \label{eq:defq}
\end{align}
with the angular frequency $\omega = 1/\sqrt{L_E(C_E - C_c^2/C_0)}$.

Now, we introduce the variable $\varphi = \frac{2e}{\hbar} \phi_0$ as well as its conjugate variable $q_{\varphi}$ satisfying $[\varphi, q_\varphi] = -i\hbar$. We define $M = (C_0 - C_c^2/C_E)/(2e)^2$ and substitute Eq. \eqref{eq:defphi} and \eqref{eq:defq} into the Hamiltonian, while keeping only energy conserving terms for the cavity field mode and ignoring constant energy shifts and get
\begin{align}
H =& \; \frac{q_\varphi^2}{2M} - E_J(\cos \varphi + I\varphi) \nonumber \\
&+(\hbar \omega + \hbar \,\eta \cos \varphi ) \, a\dag a \nonumber \\
&+\hbar \, \kappa \cos \varphi \,a\dag a\dag a a + \hbar\lambda\, q_\varphi q  \nonumber \\
&+(\hbar \mu + \hbar \chi \,a\dag a) \, \sin (\varphi - \varphi_J) \,\phi. \label{eq:hamil}
\end{align}
This Hamiltonian is the main result of this section. We recognize the Hamiltonian for a single Josephson junction and a single resonator mode coupled by linear and non-linear terms. The constants in the Hamiltonian are given as
\begin{align}
\eta &= \frac{E_J}{2} \frac{(2e)^2}{\hbar^2} \cos^2 kd \, L_E \omega \label{eq:eta}\\
\kappa &= -\frac{E_J}{4} \frac{(2e)^4}{\hbar^3} \cos^4 kd \, L_E^2 \omega^2 \\
\lambda &= -\frac{2e}{\hbar} \frac{C_c}{C_0C_E - C_c^2} \\
\mu &= -\frac{E_J}{\hbar} \frac{2e}{\hbar} \cos kd\, \cos \varphi_J \\
\chi &= \frac{E_J}{4\hbar} \frac{(2e)^3}{\hbar^2} \cos^3 kd \, L_E \omega \, \cos \varphi_J. \label{eq:chi}
\end{align}
Remembering that $\cos kd$ is assumed to be small, due to the weak field in the resonator, the magnitude of the strengths in frequency units supposedly follow the order $\mu\sqrt{\frac{\hbar \omega L_E}{2}} > \eta > \sqrt{\frac{\hbar \omega L_E}{2}}\chi > \kappa$, while $\lambda\sqrt{\frac{\hbar}{2 \omega L_E}}$ does not directly relate to the other quantities. It should also be noted that weak terms which include $\cos \varphi$ but no field coupling terms, $a$ or $a\dag$, are neglected as they merely change the Josephson energy, $E_J$, by a small amount.

\subsection{Validity of the single-mode approximation}
The Hamiltonian \eqref{eq:hamil} assumes the near-resonant coupling to only one active resonator mode. The single-mode approximation is valid when the energy difference between modes is much larger than the coupling strengths, but in superconducting circuits, $\mu\sqrt{\frac{\hbar \omega L_E}{2}}$ may be comparable to the mode frequencies, and a more careful analysis is needed.

To illuminate the discussion, we will write the Hamiltonian as
\begin{align}
H = H_{JJ} + H_{a,JJ} 
\end{align}
with $H_{JJ}$ equal to the two first terms of Eq. \eqref{eq:hamil} and $H_{a,JJ}$ equal to the rest of the terms involving the fundamental mode $a$ of the resonator. Now, we include a second resonator mode $b$, and we write the Hamiltonian 
\begin{align}
H = H_{a,JJ} + H_{JJ} + H_{b,JJ} + H_{a,b} \label{eq:Habjj}
\end{align}
with $H_{b,JJ}$ similar to $H_{a,JJ}$ and $H_{a,b}$ representing the direct coupling terms between the two modes caused by the spatially dependent terms in Eq. \eqref{eq:lagran}.

We assume that the lowest state of the coupled system is approximately a product state, 
\begin{align}
\Psi_0 = \psi_a^0 \, \psi_{JJ}^0 \, \psi_b^0
\end{align}
while the first excited states for the Hamiltonian in Eq. \eqref{eq:Habjj} may be expanded on,
\begin{align}
\Psi_1 = \psi_{a,JJ}^1 \, \psi_b^0,
\end{align}
and
\begin{align}
\tilde{\Psi}_1 = \psi_a^0 \, \psi_{b,JJ}^1,
\end{align}
where $\psi_{a,JJ}^1$ and $\psi_{b,JJ}^1$ are the first excited eigenstates of $H_{JJ} + H_{a,JJ}$ and $H_{JJ} + H_{b,JJ}$, respectively.

The validity of the single mode approximation is determined by the coupling of the ground state $\Psi_0$ and the excited state $\Psi_1$ to $\tilde{\Psi}_1$,
\begin{align}
g_{1a,1b} &= \int \tilde{\Psi}_1^* (H_{b,JJ}+H_{a,b}) \Psi_1 \\
g_{0,1b} &= \int \tilde{\Psi}_1^* (H_{b,JJ}+H_{a,b} \Psi_0.
\end{align}
The system will be driven close to the resonance between $\Psi_0$ and $\Psi_1$, and if the conditions
\begin{align}
|g_{1a,1b}| \ll |\Delta_{1a,1b}| && |g_{0,1b}| \ll |\Delta_{0,1b}| \label{eq:cond}
\end{align}
are satisfied, where we have $\Delta_{x,y}$ is the energy difference between states $x$ and $y$, the coupling can be neglected and the single-mode approximation is justified. 

In the calculations in Sec. \ref{sec:time} we find numerically the conditions in Eq. \eqref{eq:cond} to be almost satisfied. With parameters used later in the text we find $|g_{1a,1b} / \Delta_{1a,1b}| \lesssim 0.1$. This will lead to a pertubation of the energy of $\Psi_1$ by $\sim \hbar \Delta_{0,1a}/100$, and to a population in $\tilde{\Psi}_1$ of 1\%. The calculations in Secs. \ref{sec:spectral} and \ref{sec:time} are performed using the single-mode approximation, and they may hence slightly overestimate the efficiency of the device at the level of 1\%. For higher excited states and higher order resonator modes, we find a decreasing ratio between the coupling strengths and the energy differences, but further theoretical analyses into this matter will be needed to clarify the influence of more modes and levels.

\subsection{Off-resonance and multi-mode interaction}
In the previous subsection, in order to simplify the problem, we ignored multi-mode interaction from the term $\dot{\phi}\dot{\phi}_0$, which transforms into coupling of all $q_j$'s. Another way to simplify the problem is to choose the parameters of the resonator such that the energy splittings are much smaller than the energy splittings in the Josephson junction. Then we safely ignore $\phi_0$ as a dynamical variable and replace it with the static value $\phi_J$ and from \eqref{eq:defq} we get $q_j = C_j \phi_j$ with
\begin{align}
C_j = \frac{C_T d}{2} \Big( 1 + \frac{\sin 2k_j d}{2k_j d}\Big).
\end{align}
We can also use the Euler-Lagrange equation \eqref{eq:el-k}, and obtain the effective inductance
\begin{align}
L_j^{-1} = \frac{(k_j d)^2}{2L_T d} \Big( 1 + \frac{\sin 2k_j d}{2k_j d}\Big).
\end{align}
Defining $\omega_j = 1/\sqrt{L_jC_j}$ and following the derivation of Eq. \eqref{eq:hamil} we then get the Hamiltonian
\begin{align}
H = \sum_j \big( \hbar \omega_j a_j\dag a_j + \hbar \kappa_{jj} a_j\dag a_j\dag a_j a_j \Big) + \sum_{i\neq j} 2 \hbar \kappa_{ij} a_i\dag a_i a_j\dag a_j
\end{align}
with
\begin{align}
\kappa_{ij} &= -\frac{E_J}{4\hbar} \frac{(2e)^4}{\hbar^2} \cos^2 k_id \cos^2 k_id \, L_E^2 \omega^2 \cos \varphi_J.
\end{align}
This Hamiltonian is formally equivalent to the JPA Hamiltonian with $\kappa_{00}$ being the fundamental JPA Kerr non-linearity \cite{eichler2013controlling}. Unlike the usual set-up for a JPA, however, we now have an easy way of tuning the non-linearity since $\kappa_{ij} \propto \cos \varphi_J$ can be controlled by the bias current.

\section{Spectral Analysis}
\label{sec:spectral}

\begin{figure*}[t]
\includegraphics[width=1.95\columnwidth,clip=true,trim=2cm 19.6cm 2cm 3cm]{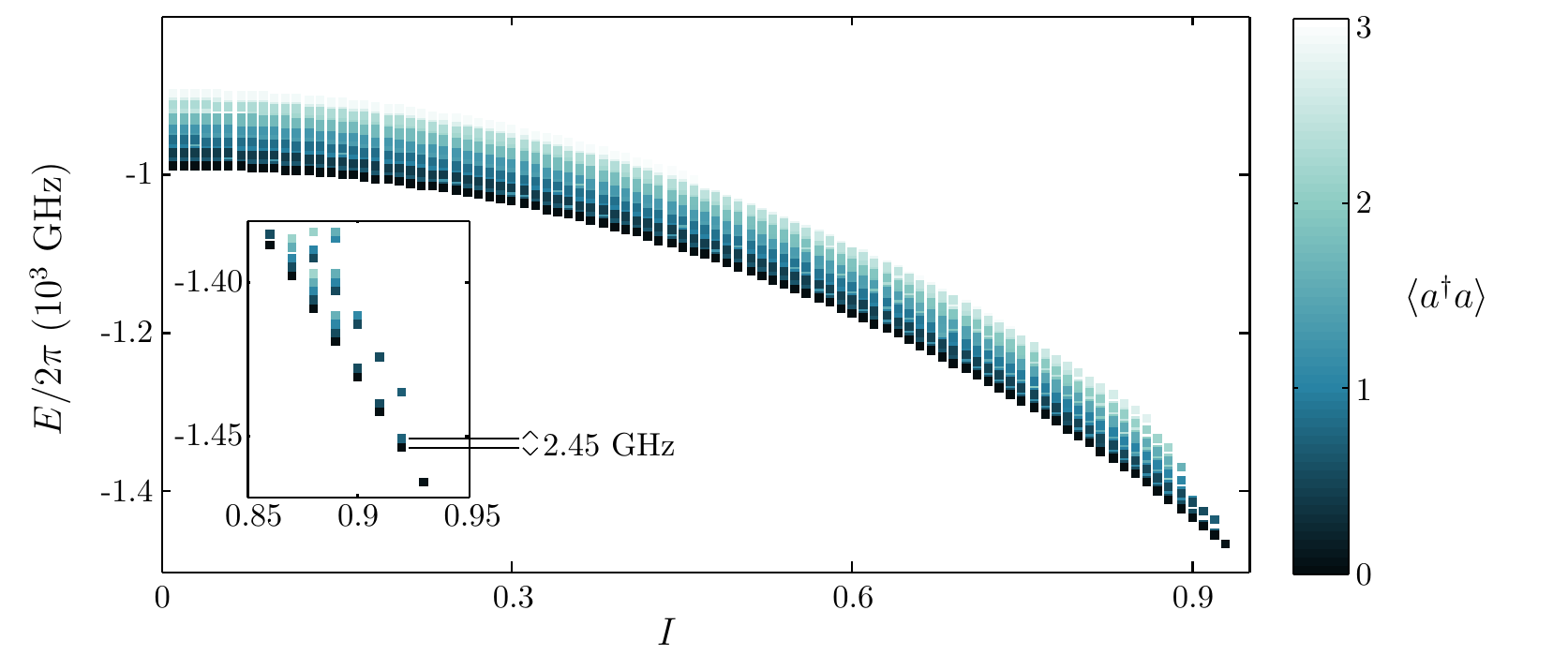}
\caption{(Color online) Eigenenergies of the Hamiltonian \eqref{eq:hamil} for different values of the dimensionless bias current $I$. The color represents the meanvalue of $a\dag a$. The parameters are chosen to represent a Josephson junction with a critical current of 2 $\mu$A and a capacitance of 1500 fF. The 50 $\Omega$-impedance resonators bare frequency is chosen to be 7 GHz. Eigensolutions with a mean occupation number in the resonator mode above 3 are not included. In the inset is a zoom of the lowest bands near the end of the bands. Marked with lines in the inset are the eigenenergies in the lowest band at $I=0.92$, highlighting the energy difference between the only two bound state of the lowest band.} \label{fig:diag_I}
\end{figure*}

\begin{figure}[t]
\includegraphics[width=0.95\columnwidth,clip=true,trim=2cm 17.7cm 5cm 2cm]{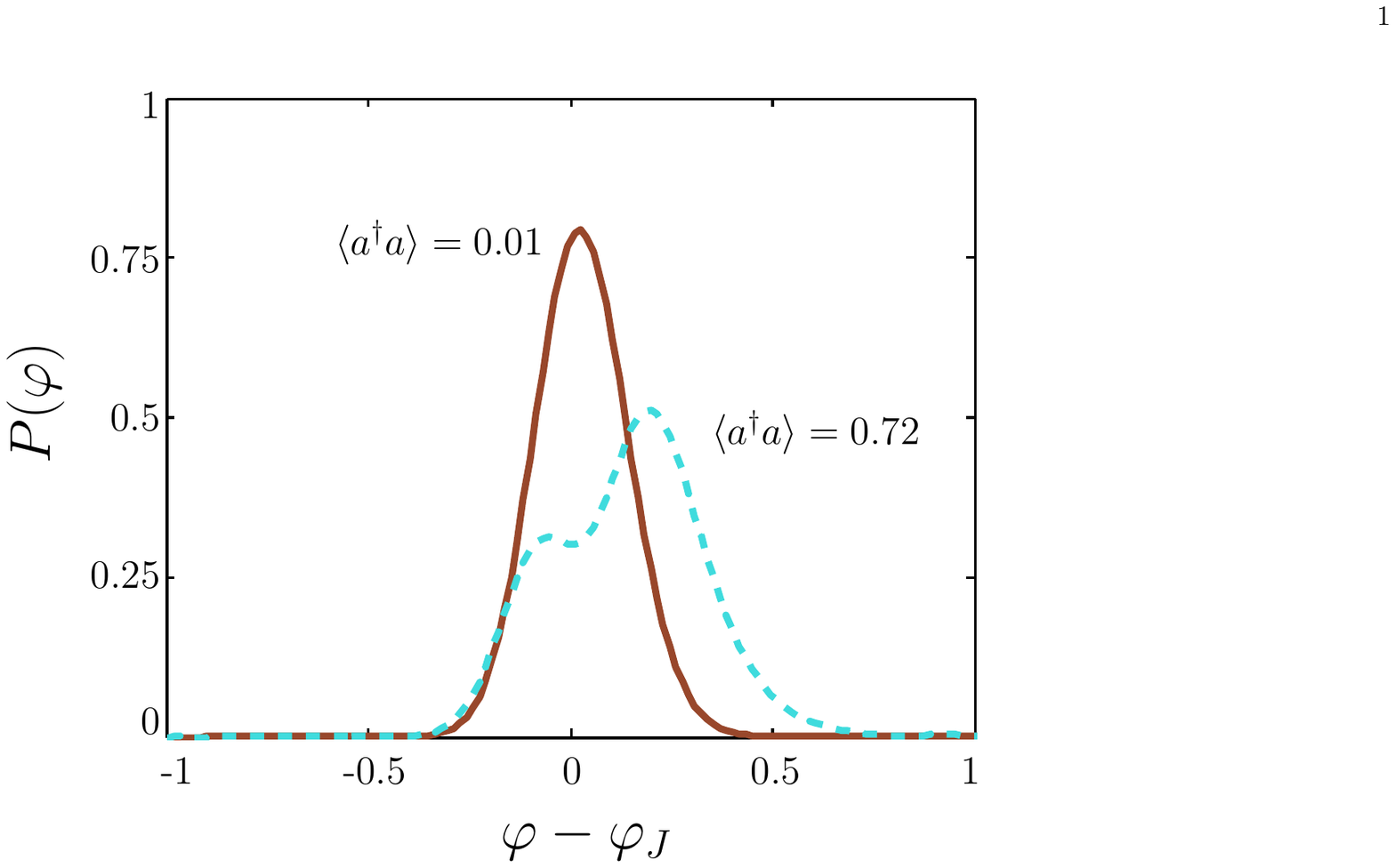}
\caption{(Color online) Probability distribution for the Josephson junction phase variable, $\varphi$, calculated with the parameters as in Fig. \ref{fig:diag_I} and $I=0.92$. This yields the following values for the parameters in Eqs. \eqref{eq:eta}-\eqref{eq:chi}, $(\eta,\kappa,\lambda\sqrt{\frac{\hbar}{2 \omega L_E}},\mu\sqrt{\frac{\hbar \omega L_E}{2}},\chi\sqrt{\frac{\hbar \omega L_E}{2}}) = (5.78,\, 0.03,\, 0.90,\, 29.7,\, 0.08)\times 2\pi$ GHz. The solid (red) line is the calculated ground state, while the dashed (blue) line is first excited state. The mean value of $a\dag a$ for the two states is also indicated in the figure.} \label{fig:wave}
\end{figure}

After having derived the single-mode Hamiltonian \eqref{eq:hamil}, we can choose the realistic parameters \cite{macha2010losses,oelsner2013underdamped,PhysRevB.81.144518}. We assume a Josephson Junction with a zero-coupling critical current $I_c = 2 \, \mu \text{A}$ and a Josephson capacitance $C_J = 1500$ fF together with a $50 \, \Omega$-impedance resonator with a bare resonance at $7$ GHz. For these parameters we get $\mu\sqrt{\frac{\hbar L_E \omega}{2}}/\omega \approx 3.5$ for $I=0.9$, which means that we have ultra-strong coupling between the two system, hence the rotation wave approximation breaks down in this regime and we cannot approximate the system by a Jaynes-Cummings-type Hamiltonian.

In Fig. \ref{fig:diag_I} we have numerically diagonalized Eq. \eqref{eq:hamil} for parameters chosen as described above. In the diagonalization procedure, the junction phase variable is described on a grid  in a box of length $2.5\pi$ leaving only one well in the potential of Eq. \eqref{eq:H_jj} and the states are now identified as localized wave-packets quasi-bound in the well. We see a band-like structure given by the number of photons in the resonator and we observe that the higher bands gradually disappear when the bias current is increased. In the end, only the empty cavity with the junction in the ground state survives as a bound state. However, even this state is not bound for $I=0.94$, which implies that the coupling to the resonator effectively changes the critical current of the junction, as one would expect.

It is also interesting to look at the wave function, $\Psi$, for the eigenstates. We expect the resonator field mode and junction phase to be highly correlated due to the ultra strong coupling terms, however we can still define the phase distribution,
\begin{align}
P(\varphi) = \int d\phi \, |\Psi(\phi,\varphi)|^2 ,
\end{align}
which we have depicted in Fig. \ref{fig:wave} for the first two eigenstates at $I=0.92$. Since the numerical calculations are done in a Fock basis for the resonator degree of freedom, we use the partial trace, $\bra{\varphi} \, \tr[res]{ \ket{\Psi}\bra{\Psi} }\, \ket{\varphi}$, to calculate the probability distribution. As an interesting feature we see that, even though the ground state is nearly symmetric, the first excited state is very asymmetric. The population in the one-photon state of the resonator effectively lowers the barrier of the potential from Eq. \eqref{eq:H_jj} [see also Eq. \eqref{eq:Ueff}], which in turn pushes the probability distribution for $\varphi$ towards the continuum.

\subsection{Coupling of the eigenstates by an external field}
We are now interested in driving transitions between the eigenstates of the system. Such a coupling can be realized by coupling the system to the field in an outside resonator through a capacitor with the capacitance $C_{out}$ at $x=0$ (see dashed box in Fig. \ref{fig:circuit}).

Adding the capacitor gives rise to a Lagrangian term
\begin{align}
\mathcal{L}_{out} = \frac{C_{out}}{2} \big(\dot{\phi}_{out} - \dot{\phi}(0) \big)^2
\end{align}
which yields terms quadratic in both $\dot{\phi}(0)$ and $\dot{\phi}_{out}$, but typically $C_{out}$ is much smaller than any other capacitive element so we neglect these terms. In this approximation the canonical variables are not changed. We can therefore write
\begin{align}
\dot{\phi}(0) &= \frac{C_0 - C_c}{2(C_c^2 - C_EC_0)} q + \frac{C_E - C_c}{2(C_c^2 - C_E C_0)} q_0 \\
\dot{\phi}_{out} &= \sqrt{\frac{\hbar\, \omega_{out}^2 Z_{out}}{2}} (b + b\dag),
\end{align}
with $Z_{out}$ being the impedance of the outside resonator and $\omega_{out}$ its frequency. Now $b$ ($b\dag$) annihilates (creates) a photon in the the outside resonator.

In order to estimate the coupling strength of the coupling between the device and the outside field we look at the spectra from the Hamiltonian and choose a value of $I$ such that we only have two states in the lowest band. We will now denote these as $\ket{0}$ and $\ket{1}$. From this, we get a term for the Hamiltonian
\begin{align}
H_{out} = \Omega  \ket{0}\bra{1} (b + b\dag) + \text{H.c.} \label{eq:Hout}
\end{align}
with the coupling strength
\begin{align}
\Omega =& \, \alpha \, \big(\beta_1 \bra{0}q_{\varphi}\ket{1} +  \beta_2 \bra{0}a + a\dag\ket{1} \big)
\end{align}
and the quantities defined by
\begin{align}
\alpha =&\, C_{out} \sqrt{\frac{\hbar\, \omega_{out}^2 Z_{out}}{2}} \\
\beta_1 =&\, 2e\frac{C_E - C_c}{2(C_c^2 - C_E C_0)} \\
\beta_2 =&\, \sqrt{\frac{\hbar}{2L_E \omega}} \frac{C_0 - C_c}{2(C_c^2 - C_EC_0)}.
\end{align}
In writing Eq. \eqref{eq:Hout} we have neglected coupling to higher bands as well as coupling to unbound states, but if we choose $\omega_{out}$ to be resonant with the splitting in the lowest band, this should be a good approximation. Now, if we use the same parameters as before and we set $C_{out} = 5$ fF at a bias current $I=0.92$ with a frequency of the outside field resonant with the energy-spltting, $\hbar \omega_{out} = E_1 - E_0$, we get a coupling strength of $|\Omega| = 2\pi \times 29$ MHz. This coupling will also mediate a decay from $\ket{1}$ assuming no external field is applied with a time-scale set by the coupling strength.

\begin{figure}[t]
\includegraphics[width=0.95\columnwidth,clip=true,trim=2cm 17.7cm 5cm 2cm]{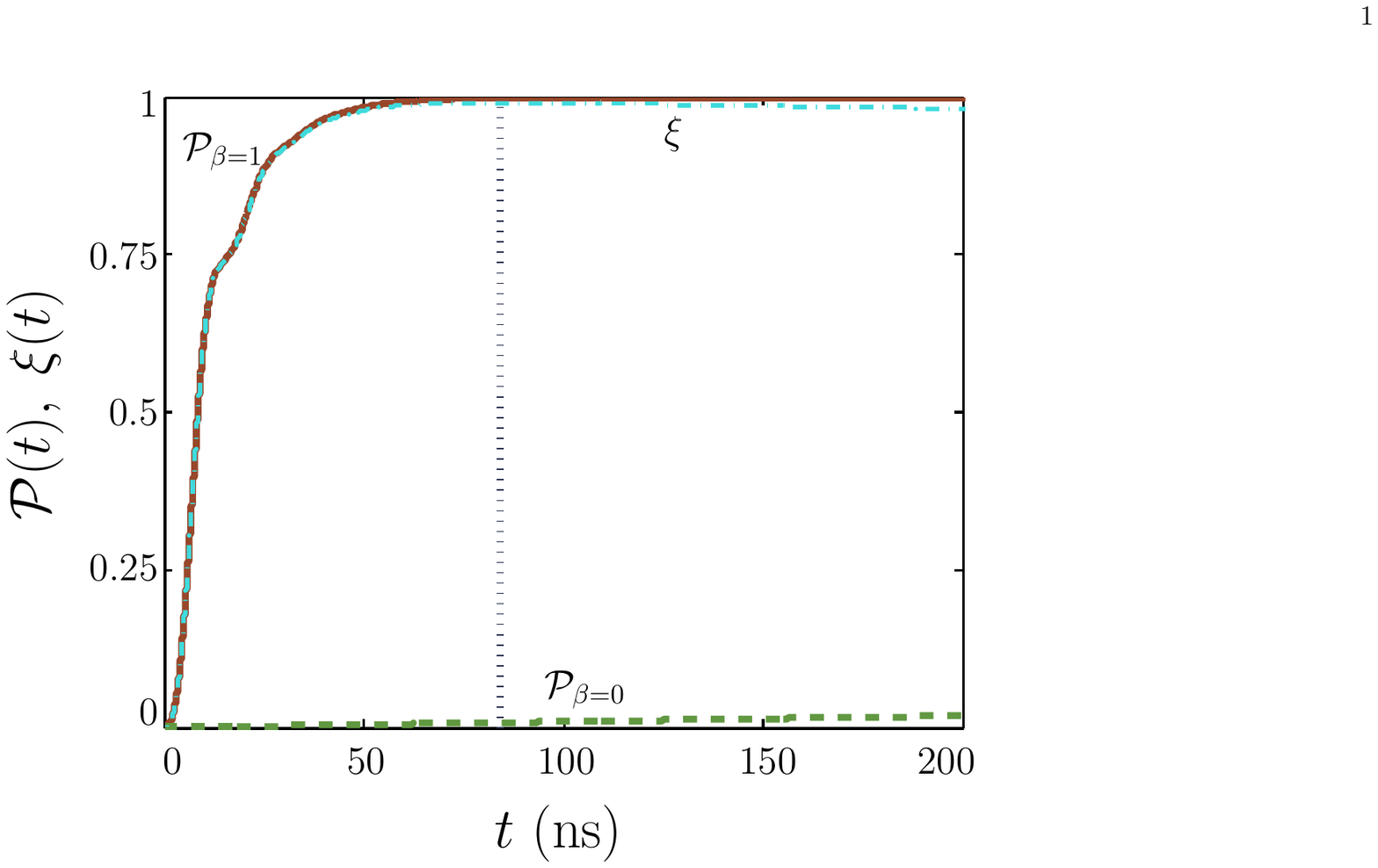}
\caption{Accumulated probability for a switching event and the detector efficiency for $I=0.92$. The switching  probability for $\beta=1$ is shown as the solid (red) line, and for $\beta=0$ as the dashed (green) line. The detector efficiency is shown as the dashed-dotted (blue) line. The dotted vertical line marks the maximal efficiency point. The parameters chosen are those of a Josephson junction with a critical current at 2 $\mu$A, a capacitance at 1500 fF and a Josephson resistance at $300\; \Omega$, thus Eqs. \eqref{eq:eta}-\eqref{eq:chi} yield the same numerical values as in Fig. \ref{fig:wave}. The 50 $\Omega$-impedance resonator bare frequency is 7 GHz, and we assume $C_{out} = 5$ fF.} \label{fig:time_P_eff}
\end{figure}

\begin{figure*}[t]

\begin{minipage}[t]{0.32\linewidth}
\hspace{0.3cm}(a)\\\includegraphics[width=0.99\columnwidth,clip=true,trim=2cm 17.7cm 6.2cm 1.43cm]{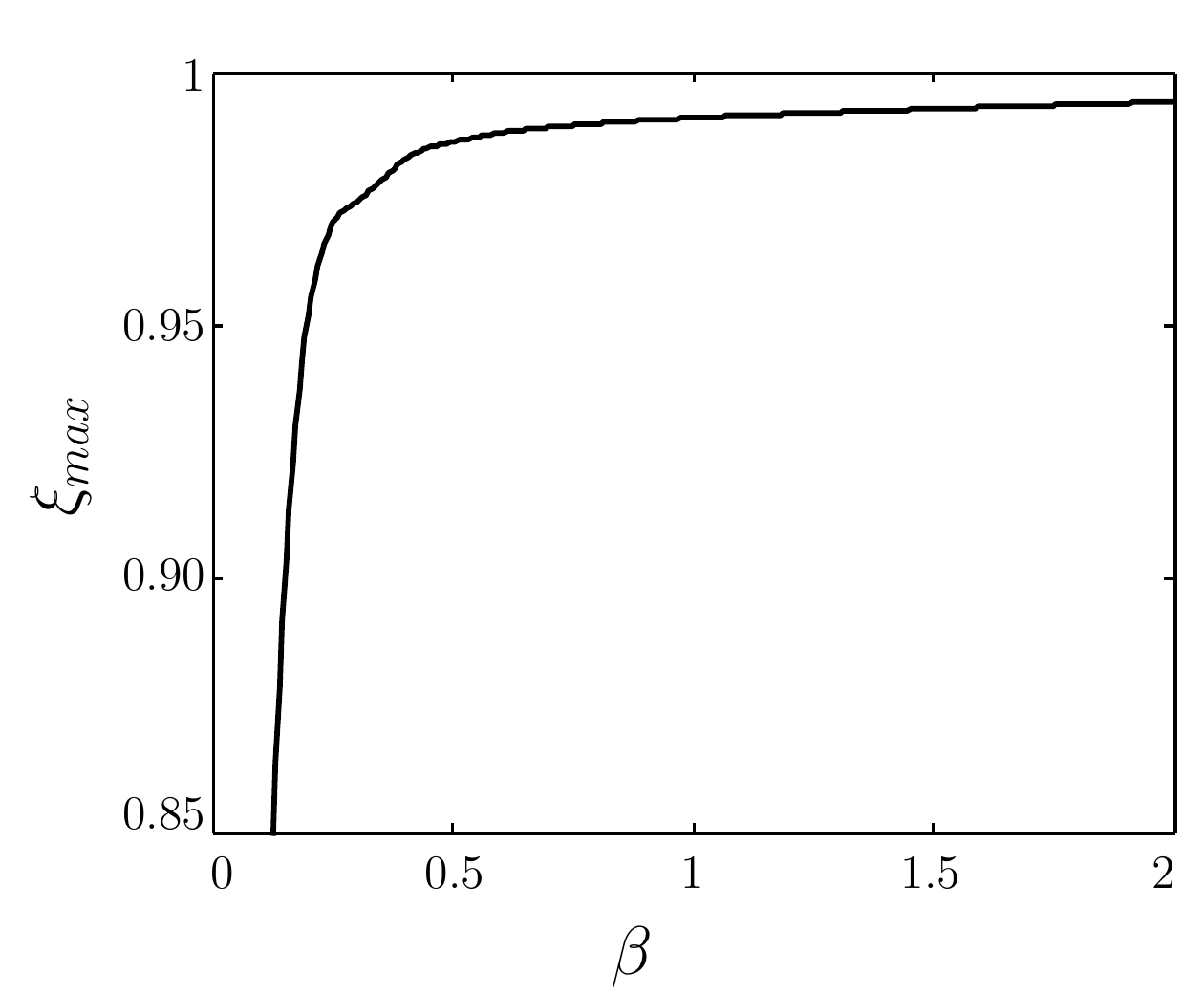}
\end{minipage}
\begin{minipage}[t]{0.32\linewidth}
\hspace{0.8cm}(b)\\\includegraphics[width=0.99\columnwidth,clip=true,trim=2cm 17.7cm 6.2cm 1.43cm]{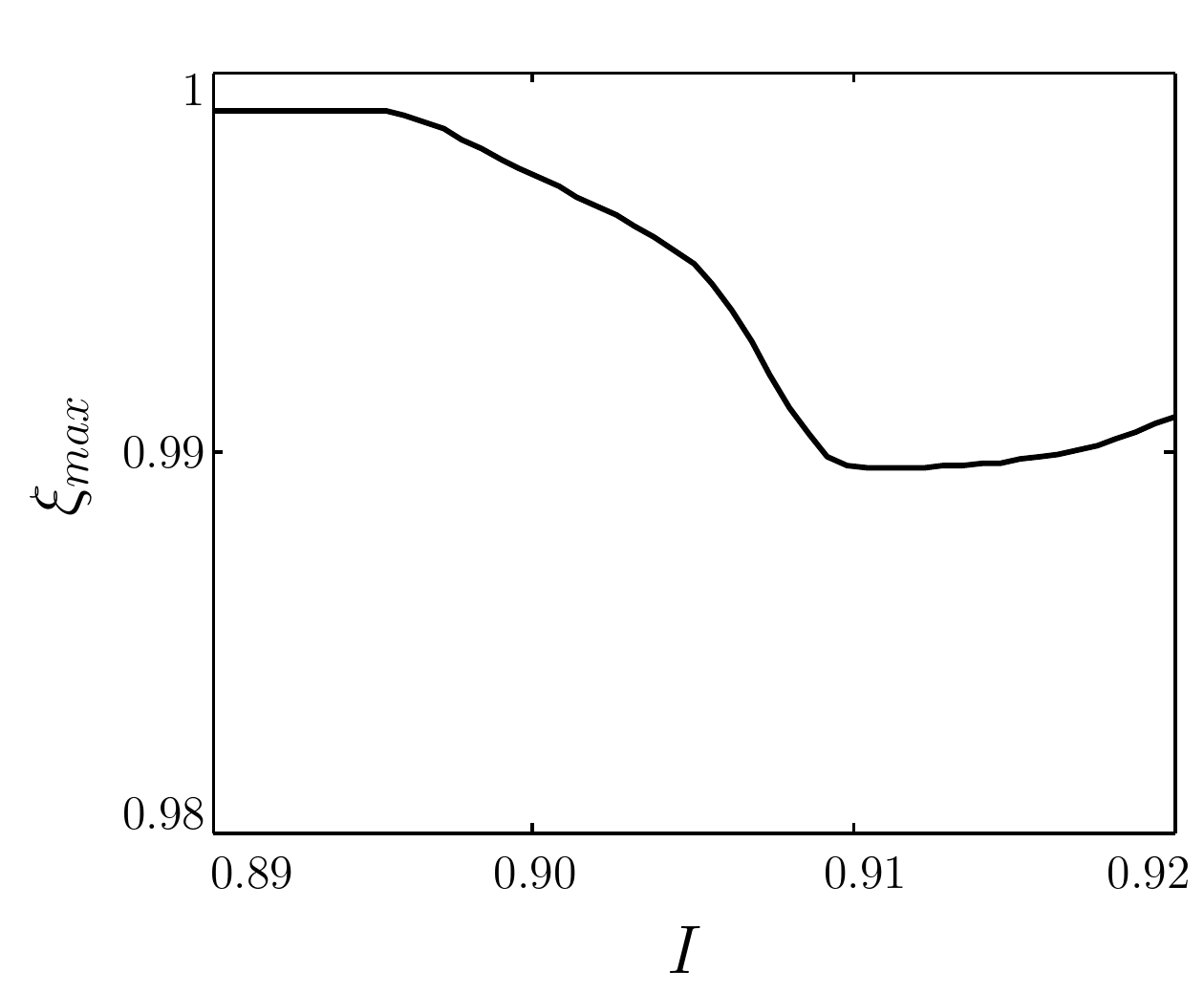}
\end{minipage}
\begin{minipage}[t]{0.32\linewidth}
\hspace{0.75cm}(c)\\\includegraphics[width=0.99\columnwidth,clip=true,trim=2cm 17.7cm 6.2cm 1.43cm]{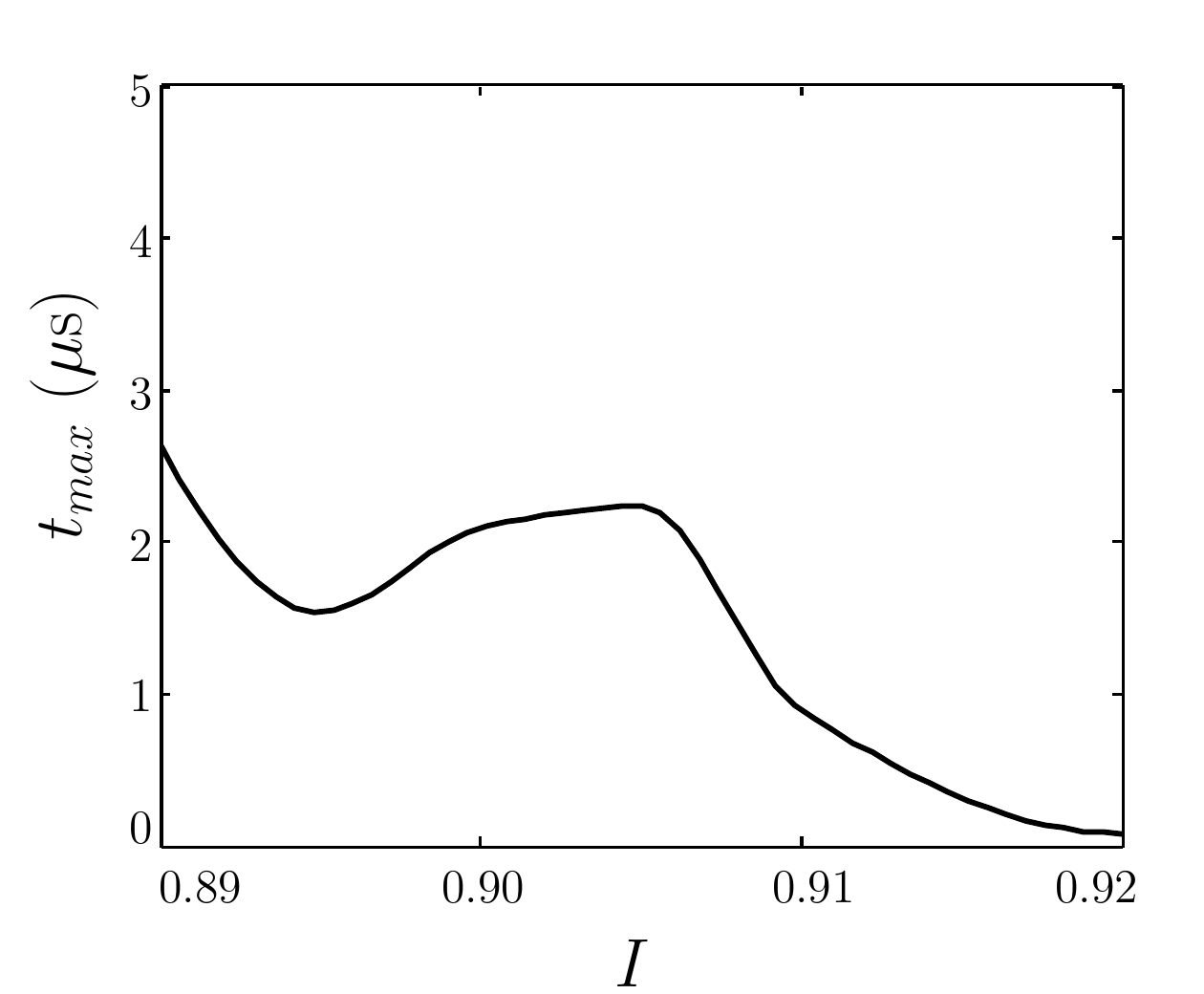}
\end{minipage}

\caption{In (a) we show the maximal detector efficiency for $I=0.92$ as a function of $\beta$, while in (b) the maximal detector efficiency for $\beta=1$ is shown as a function of $I$. In (c) we display the detection time required to reach the efficiency in (b). In all figures $\omega_{out}$ is equal to the energy splitting between the two lowest bound states. The rest of the parameters are the same as Fig. \ref{fig:time_P_eff}.} \label{fig:more}
\end{figure*}

\begin{figure}[t]

\hspace{0.8cm}(a)\vspace{-0.3cm}\\
\quad\includegraphics[width=0.66\columnwidth,clip=true,trim=2cm 17.7cm 6.2cm 1.43cm]{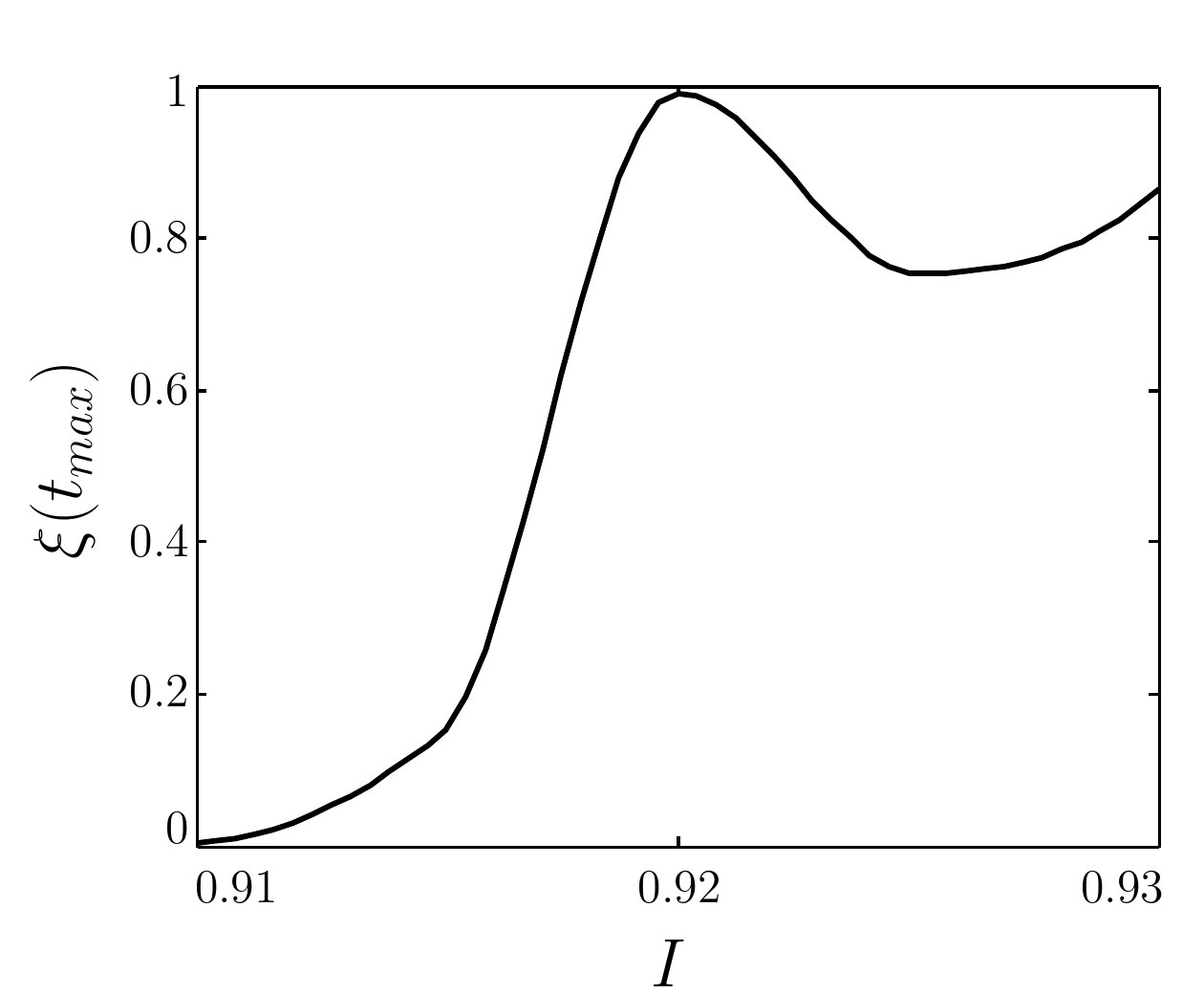}
\vspace{0.2cm}

\hspace{0.8cm}(b)\vspace{-0.3cm}\\
\includegraphics[width=0.66\columnwidth,clip=true,trim=2cm 17.7cm 6.2cm 1.43cm]{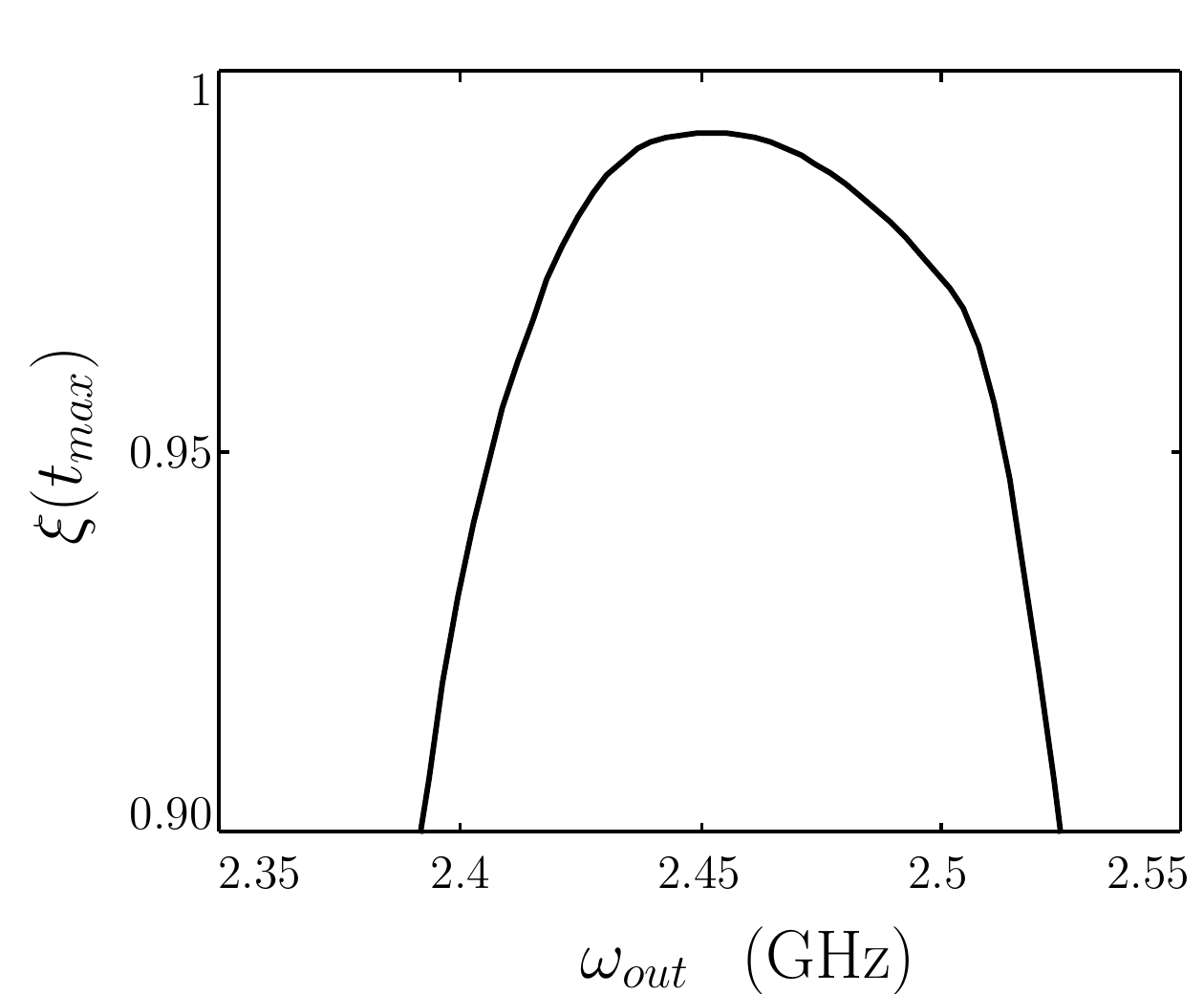}

\caption{(Color online) In (a) we show the detector efficiency as a function of $I$ for fixed $\omega_{out} = 2.45$ GHz, which is the energy-splitting between the two bound states at $I=0.92$. In (b) the maximal detector efficiency is plotted as a function of $\omega_{out}$ for fixed $I=0.92$. All efficiencies are evaluated at the optimal time, $t_{max} = 84$ ns and other parameters are once again the same as in Fig. \ref{fig:time_P_eff}.} \label{fig:920}
\end{figure}

\section{Time-dependent analysis}
\label{sec:time}

With a time-independent description in place, we now have turn to the question of time propagation of the system. If the system is prepared in an eigenstate of the Hamiltonian the time evolution is at a first glance trivial, however if the system is prepared with a bias current close to the critical current, there is a finite chance of tunneling through the potential barrier causing a voltage switch across the junction  \cite{PhysRevB.81.144518,PhysRevA.87.052119,PhysRevB.72.134528,PhysRevB.68.060504,PhysRevB.77.104531}.

Following the description in Ref. \onlinecite{PhysRevA.87.052119}, we treat the tunneling loss process by propagating the wave function of the phase variable in a time-dependent imaginary potential (TDIP),
\begin{align}
i\hbar \pfrac{\Psi}{t}  = (H -iV_{im}(t) )\Psi.
\end{align}
The method of \cite{PhysRevA.87.052119} also includes a Markovian friction term to take the junction resistance into account. In the simulations we have used a Josephson resistance at $300\; \Omega$.

We use here an ansatz for the TDIP evaluated at each time-step as a function of the resonator field mode variable, $\phi$. Taking the mean of resonator operators in each time-step provides an effective potential for the phase particle
\begin{align}
U_{eff}(\varphi) =& \,\hbar \omega \bar{n} - E_J\Big( \cos \varphi \big(1 + \frac{\hbar \eta}{E_J} \bar{n} + \frac{\hbar \kappa}{E_J} \exv{a\dag a\dag a a}\big) \nonumber\\ &\phantom{\hbar \omega \bar{n} - E_J\Big(} +\, \varphi \big( I + \frac{\hbar \mu}{E_J} \exv{\phi} + \frac{\hbar \chi}{E_J} \bar{n} \exv{\phi} \big) \Big), \label{eq:Ueff}
\end{align}
with $\bar{n} = \exv{a\dag a}$ being the mean photon number in the resonator field mode at a given time. By taking the mean values we neglect an amount of correlations between the tunneled phase-particle and the resonator, however the tunneling and detection time is much faster than the characteristic time scale of these correlations. With this potential, we can determine the time dependent classical turning point and, following  \cite{PhysRevA.87.052119}, a useful expression for $V_{im}(\varphi)$.

%\subsection{Constant driving field}

In the following we assume that our device is initialized in the ground state and that the resonator is driven with a constant weak classical field, so that
\begin{align}
(b + b\dag)(t) \rightarrow \beta \, \sin \omega_{out} t
\end{align}
with $\beta$ a constant of order $1$ \cite{PhysRevA.77.052111,carmichael2008statistical}. We thus neglect the operator character of the incident field and the decay from the resonator mode into field modes outside the resonator. Equation \eqref{eq:Hout} then yields
\begin{align}
H_{out}(t) = \alpha \beta \Big( \beta_1 q_{\varphi} + \beta_2 (a + a\dag) \Big)\sin \omega_{out} t.
\end{align}

In Fig. \ref{fig:time_P_eff} we present the calculated probability that a switching event has happened, calculated as
\begin{align}
\mathcal{P}(t) &= 1 - ||\Psi(t)||^2 \label{eq:prob}
\end{align}
evaluated at the time $t$ and we define the detector efficiency as in  \cite{PhysRevB.86.174506}, $\xi(t) = \mathcal{P}_{\beta = 1}(t) - \mathcal{P}_{\beta = 0}(t)$. The norm $||\Psi(t)||^2$ is expected to decrease due to the propagation in the TDIP and the tunneling rate can be calculated from the derivative $\gamma = -d||\Psi(t)||^2/dt$. We see that for $\beta = 1$ we approach unit probability within roughly 80 ns, which we may compare with the Rabi time $t_r = \pi/|\Omega| \approx 21$ ns, which implies that around 4 Rabi oscillations are made before a tunneling event is certain. We recall, however, that the Rabi oscillations are modified due to the non-linear nature of the detector. Nevertheless, the first oscillation can be observed in Fig. \ref{fig:time_P_eff} as a shoulder on the probability graph around half the Rabi time.

We have also marked the most efficient point in Fig. \ref{fig:time_P_eff}, that is the maximum of $\xi(t)$. We will denote this efficiency $\xi_{max}$ and the time where it occurs $t_{max}$. For the parameters in Fig. \ref{fig:time_P_eff} we get $\xi_{max} = 0.991$ with $t_{max} = 82$ ns. Figure \ref{fig:more} now shows the maximal efficiency as a function of both $I$ and $\beta$. In each case the driving frequency, $\omega_{out}$, is equal to the energy splitting of the two lowest bound states. This restricts these simulations to $I\leq 0.92$, as we do not have more than one bound state above this bias-current. As expected, the maximal efficiency increases as the field strength, $\beta$, increases as seen in Fig. \ref{fig:more} (a). However, once we are above $\beta=0.5$, $\xi_{max}$ saturates. Note that, in our description, a change in $\beta$ is equivalent to a change in $C_{out}$. In Fig. \ref{fig:more} (b) we see that changing the current to a lower bias-current opens for the possibility of even higher quantum efficiency, but in Fig. \ref{fig:more} (c) we see that it comes at a price of significantly larger detection time. For the large detection times required for $I<0.91$, we might not be able to safely neglect decay in the resonator as we have done in these calculations, thus the efficiency for these values may be smaller than shown in Fig. \ref{fig:more} (b). To summarize, we see that our efficiency is close to unity when $\beta > 0.5$ and we get the shortest detection time when $I=0.92$.

In Fig. \ref{fig:920}, we characterize the performance of the detector for a detection time at 82 ns. In Fig. \ref{fig:920}(a), we notice that for $I<0.92$ we quickly lose performance, while at larger $I$ we retain a good detection efficiency. This we can interpret as at lower $I$ we get a suppressed decay rate due to the narrow linewidth of the second lowest energy state in the device, while at larger $I$ we excite directly into the continuum, since only one bound state is present. Transferring population directly into the continuum is a weaker process than resonant transfer via an excited bound state, but still stronger than going via a far-detuned narrow state \cite{PhysRevA.55.1262,PhysRevLett.65.104,PhysRevLett.67.516}. Finally, a slight increase in the efficiency is observed at $I=0.93$, but here the ground state is very unstable and if the bias current is increased further, no bound state is present in the device.

Figure \ref{fig:920}(b) shows that, we have a frequency band of around 100 MHz with efficiencies above 0.9, which is substantially larger than the linewidths of  state-of-the-art resonators and qubits in cQED \cite{niemczyk2010circuit}. The device thus offers adequate detection efficiency of microwave signals from cQED experiments. We can estimate the relaxation time $T_1 = 1/\Delta \omega$, with $\Delta \omega$ the full width at half maximum of a Lorentzian fit \cite{PhysRevB.68.060502}. This yields $T_1 \approx 7.5$ ns, which is a typical order of magnitude for this type of phase qubits \cite{PhysRevB.86.174506,PhysRevB.68.060502}. It is worth noting that $T_1$ can be optimized by design of the qubit to improve performance on resonance \cite{PhysRevB.86.174506}, however at the expense of a limited bandwidth of the detector. The value of $T_1$ is smaller than the Rabi-time, but this fact does not limit the performance significantly. Coupling the CBJJ directly to the $\lambda/4$ resonator mitigates the limitations imposed by a small $T_1$ in the setup proposed in \cite{PhysRevB.86.174506}.

\section{Conclusion and Outlook}
\label{sec:conclusion}

In this paper, we have derived the Hamiltonian for a $\lambda$/4-resonator shunted by a current-biased Josephson junction (CBJJ). This device was expected to work as a very sensitive microwave detector near the quantum limit, as it combines the techniques of a JPA to amplify the incoming signal, with the voltage switch of a CBJJ to detect the signal. Numerical calculations show that we indeed get a very high detector efficiency of the device.

Using recently developed theory to describe the switching of a CBJJ \cite{PhysRevA.87.052119}, the calculations take into account both the complex tunneling dynamics of a CBJJ as well as relaxation in the junction, however we have neglected losses in the resonator. The method to describe the tunneling uses a time-dependent imaginary potential (TDIP), which is shown in \cite{PhysRevA.87.052119} to be a good approximate method. We derive the Hamiltonian using a standard method for quantization of electric circuits \cite{devoret1995quantum,PhysRevA.29.1419} and we get a coupled resonator-like degree of freedom and CBJJ-like degree of freedom. This allows us to use the method of \cite{PhysRevA.87.052119} to describe the tunneling. We emphasize that we have extended the model of  \cite{PhysRevA.87.052119} to a junction coupled to a quantized field, but we evaluate the TDIP using mean values of the field. This is an approximation that assumes fast detection of the tunneled phase.

Furthermore, to make sure that resonator losses may be neglected, we seek short optimal detection time, which we get by going to the highest bias current, $I=0.92$, where two bound states still remain in the full system. Here, we get a quantum efficiency at $0.991$ at a detection time of 82 ns. Tuning the bias current to a higher value will reduce the efficiency, as bound excited states are lost. Finally we have shown that at the optimal bias-current, within a frequency bandwidth of approximately 100 MHz the efficiency is above 0.9. The calculations have been done based on a single-mode approximation, thus an experimental implementation might suffer from a small leakage into higher modes of the resonator and thus the efficiency will be slightly reduced or a slightly longer detection time will be required.

The device may be built with current technology, and experimental implementations will provide further insight to the dynamics and performance of the device. Of special interest from a quantum information point of view is the dynamics when the device is coupled to one or more qubits, as the measurement back-action from quantum measurements is known to lead to non-trivial evolution of qubits  \cite{murch2013observing,PhysRevLett.109.050506,PhysRevLett.109.050507}. The measurement back action on the field degree of freedom of the device is considerably different from the application of the annihilation operator in conventional photon detection \cite{PhysRev.130.2529}, and the use of the CBJJ for quantum field detection may thus stimulate development of a novel quantum measurement theory in the microwave domain.

\section*{Acknowledgements}
\label{sec:ack}
We thank A. C. J. Wade for a careful reading of the manuscript. The authors acknowledge support from the EU 7th Framework Programme collaborative project iQIT.

\bibliography{bt}

\end{document}